# An Era of Precision Astrophysics: Connecting Stars, Galaxies and the Universe
## an Astro2010 Science White Paper

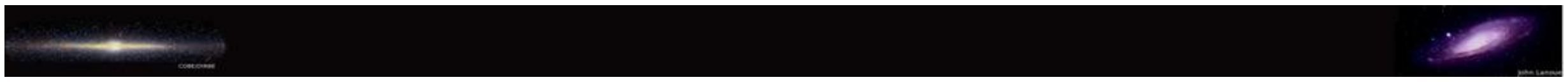

Rob P. Olling[UMD],
Ron J. Allen[ST], Jay Anderson[ST], Brian C. Chaboyer[DC], Wendy Freedman[CO],
Puragra Guhathakurta[UCSC], Kenneth Johnston[USNO], Shri Kulkarni[CALT],
Sebastien Lepine[AMNH], Valeri V. Makarov[NC], Eric E. Mamajek[UR],
Alice C. Quillen[UR], Kailash S. Sahu[ST], Ata Sarajedini[UF], Ed J. Shaya[UMD],
Donald Terndrup[OS], Patrick A. Young[ASU]

[AMNH]American Museum of Natural History, NY
[ASU]Arizona State University
[CALT]Caltech
[CO]Carnegie Observatory
[DC]Dartmouth College
[NC]NexScI, Caltech
[OS]Ohio State University
[ST]Space Telescope Science Institute
[UCSC]UC, Santa Cruz
[UF]University of Florida
[UMD]University of Maryland
[UR]University of Rochester
[USNO]U.S. Naval Observatory, Washington DC

## ABSTRACT

The golden age of astrophysics is upon us with both grand discoveries (extra-solar planets, dark matter, dark energy) and precision cosmology. Here we argue that fundamental understanding of the working of stars and galaxies is within reach, thanks to the precision measurements that are now becoming possible. We highlight the importance of micro-arcsecond astrometry which forms the basis of model independent distances and masses.

Distances are one of the most fundamental external properties of astronomical objects and accurate knowledge of distances can change our perception of astronomical phenomena dramatically. For example, in antiquity, the idea that the Earth revolved around the Sun was rejected because the predicted annual parallax of stars were not observed. In modern times, just before the first distance measure of gamma-ray bursts, there were more theories about the nature of these objects than the number of observed $\gamma$-ray bursts.

Some of the strongest motivations to vigorously pursue accurate distance measurements are related to the history and fate of the universe. Two of the three methods available to date the universe are based on accurate distance measurements. The first method is based on the ages of stars, which can only be ascertained if their luminosities (and hence distances) are accurately known. The second method relies on cosmological methods. To first order, the age of the universe is the inverse of Hubble's constant ($H_0$), while further corrections depend on the fate of the universe -closed, critical or open- and on the amount and nature of dark energy.

Indeed, the previous decadal report stresses that: “the fundamental goal of ... astrophysics is to understand how the universe ... galaxies [and] stars ... formed, how they evolved, and what their destiny will be” (McKee & Taylor, 2001). These age-old questions can be answered to a large extent by science topics that rely significantly on micro-arcsecond astrometry: 1) Galactic archeology: a detailed reconstruction of the formation history of the Milky Way and other Local Group galaxies, 2) the very oldest stars in the Milky Way and the age of the Universe, and 3) $H_0$ and concordance cosmology. *In question form we can summarize these goals as: 1) What is the construction history of the Milky Way and other nearly galaxies? 2) what is the age, density and curvature of the Universe?*

*These goals are achievable in the near future by combining micro-arcsecond astrometry from the proposed SIM-Lite mission (Unwin et al. 2008) with data from a modest ground-based spectroscopic observing campaign. The high-quality data like we are advocating for in this white paper will force the biggest reassessment of stellar astrophysics in more than 50 years, and its effects will be very beneficial for many disciplines of astrophysics.*

## INTRODUCTION

The importance of distance determination in astrophysics is reflected by the myriad of methods available to do so: *trigonometric parallax, orbital parallax, secular parallax, moving clusters, eclipsing binaries, Baade-Wesselink method,* Pop-I Cepheids, dwarf Cepheids, RR Lyrae, red-clump giants, tip of the red-giant branch, Miras, *rotational parallax,* planetary-nebula luminosity function, *mega masers,* surface-brightness fluctuations, Tully-Fisher relation, Faber-Jackson relation, Hubble flow, supernova Ia, to name just a few. In the "DISTANCE MEASURES" section below, we separate the methods that depend on intrinsic properties of the objects from those methods that depend mostly on the properties of the detection apparatus, and are thus extrinsic to the sources. *The extrinsic techniques -italic* in the enumeration above- tend to be most reliable.

Distances relate directly to the astrophysical processes we are trying to understand: 1) if an object is twice as far away, it needs to produce four times as much energy (if unbeamed), and the energy-generation rates sometimes set firm limits on the possible processes, 2) a luminosity uncertainty for a star is equivalent to an age uncertainty, and 3) to first order the age of the universe is the inverse of the Hubble constant so that errors on $H_0$ lead to an uncertain age. Based on the WMAP and other data Komatsu et al. (2008) estimate the age of the universe with an accuracy of 0.9%, which will become significantly better with new data such as *Planck*.

**STARS:** As we mentioned in the Abstract, astrophysics will be entering a new era dominated by highly precise data and accurate inferences. A subset of stars known as the double-lined variety of detached eclipsing binaries (DEBs) will be particularly useful. For DEBs within 1.5 kpc or so, SIM-Lite can measure distances to better than ½%, which makes it possible to determine their ages with similar accuracies as can be accomplished with cosmological methods[1]. This "d½%-sample" would be the "gold standard" for stellar astrophysics: mass, radius, composition, temperature and luminosity will be accurately known for its members. Only data of this quality can seriously challenge models of stellar structure and evolution (e.g., Andersen, J. 1991, 2002; Lastennet & Valls–Gabaud 2002; Fernandez et al. 2002). Currently, we know the fundamental parameters of only about one hundred stars that cover a relatively small range in mass, temperature and metallicity, and so it is not surprising that there are many limitations in our current understanding of stars (e.g., Kurucz 2002; Terndrup 2008), especially at low metallicity and for high- and low-mass stars (Casagrande et al. 2007; Benedict et al. 2009). We anticipate that the stellar models will be significantly more accurate after this serious confrontation with the GAIA (Perryman et al. 2002) and SIM-Lite astrometric data sets of unprecedented size and accuracy. *The models that survive this confrontation will then serve as calibrators that allow for an accurate translation of observed spectroscopic and photometric indices into effective temperature ($T_{EFF}$), surface gravity (g), metallicity, extinction and mass (M).* While such accuracy is beyond the current state of the art (e.g., Soubiran et al. 1998; Kurucz 2002; Bertone et al. 2008), the requirements are not unreasonable or beyond reach.

*A very exciting by-product of this re-calibration of stellar models lies in the ages of stars. In fact, we expect that it will be possible to determine ages to within a percent or so for those stars that are just a bit younger than the Universe*[2].

**GALAXIES:** with micro-arcsec astrometry we can also determine the distance to the spiral galaxies in the Local Group with an accuracy of ½ - 2% via the “rotational parallax” method (Olling & Peterson 2000; Olling 2007). This rotational-parallax method is based on the fact that, loosely speaking, the distance (D) of the galaxy equals the ratio of rotation curve of the galaxy obtained from radial velocities ($V_R$) and the proper-motion rotation curve ($\mu_R$): that is to say, D $\propto V_R/\mu_R$. *The rotational-parallax method is possibly the most accurate geometric method for extra-galactic distance determination* (see Olling 2007 for a review). M31 and M33 are the most promising candidates for an accurate application of the rotational-parallax method[3].

---

1 Detached eclipsing binaries make up about 0.8% of the general stellar population, and GAIA will determine masses with an accuracy of less than 1% for ~100,000 DEBs (Wilkinson et al. 2005). *Double-lined systems make up ~¼ of the total number of DEBs: these systems allow accurate determination of the physical properties of both stars.* We expect that many of our targets would be known before SIM-Lite flies based on Pan-STARRS, on-going planetary transit searches and GAIA's early-release catalog.

2 Bertelli's isochrones indicate that a 0.8 $M_{\odot}$ low-metallicity star evolves at ~8.7% per Gyr. Thus, a distance error of ½ % (1% in luminosity), leads to an age error of roughly 1%/(8.7%/Gyr) ~115 Myr (1% error for an age of 12 Gyr).

**<u>THE UNIVERSE:</u>** The importance of an accurate geometric determination of the distance of M31 and M33 lies in the fact that it would provide a luminosity calibration of all the steps of the distance ladder that will be entirely independent of GAIA's calibration of their Galactic cousins (RR Lyrae, Cepheids, tip of the red-giant branch, clump giants, PN luminosity function, etc.). Furthermore, since all stars in M31/M33 will have distances known to about 1%, one can easily preform consistency checks by inter-comparing the various “standard candles” and investigate whether or not there are metallicity-induced regional differences. Geometric rotational-parallax distances for M31/M33 will also provide an absolute calibration for supernovae of all kinds that are bound to occur in these galaxies. Thus, rotational-parallax distances provide a geometric, and entirely luminosity-independent zero point for the Hubble constant. Thus, *zero points based on geometric distances for three or four galaxies (MW, M31, M33, LMC?), make possible a careful assessment of any possible systematic errors, rather than just the precision of the internal errors of the calibrations.*

Due to the $H_0^2$-dependence of the critical density, $H_0$ enters virtually all cosmological inferences (the ubiquitous “$h^2$” terms). Thus when future cosmological experiments such as *Planck* provide percent-level measurements, the $H_0$ errors need to be of similar order. While $H_0$ and the age of the universe can be determined by combining several different methods (e.g., Spergel et al. 2007; Komatsu et al. 2008), reliance on concordance cosmology may be risky since it has few built-in consistency checks.

*Stellar ages and an independent measure of the Hubble constant as enabled by micro-arcsec astrometry provide two independent consistency checks for concordance cosmology.*

## <u>SCIENCE APPLICATIONS & EXPECTED RESULTS</u>

**<u>STARS & MILKY WAY:</u>** SIM-Lite is absolutely essential for the creation of the “d½% sample.” So as to determine the properties of the oldest halo stars in the Milky Way, we need of order 1,000 double-lined detached eclipsing binaries with distance accuracy <~ ½%. Given the space density of G-type halo stars, one-half of these binary systems will be found between 1,100 and 1,400 pc ($m_v$ in [14.6, 15.1]) for which SIM-Lite will get errors of ~½%. This sample would cover a wide range in metallicity, mass and orbital separation. The latter to be able to assess the effects of tidal effects on the internal structure and evolution.

3 The LMC is not very suitable for this method because the random motion are large as compared to the rotational velocity, while the LMC is dynamically fairly complex. Nevertheless, GAIA data for millions of stars might be used to obtain an accurate RP distance for the LMC.

The metallicity of stars plays an important role because lower [Fe/H] means faster evolution. At [Fe/H]=-1.5, the 0.8 $M_\odot$ star takes 12 Gyr to reach then end of the main sequence, while its solar metallicity cousin takes about 20 Gyr to do so. Thus, at 0.8 $M_\odot$, $\delta$Age/$\delta$[Fe/H] ~ 8 / (-1.5) = -5.3 Gyr/dex. Current *random* metallicity errors are ~0.03 dex (Terndrup 2009), which correspond to an age uncertainty of 0.16 Gyr, or about 1.3% at an age of 12 Gyr. However, to achieve reliable ages, we need small *external* errors, which are currently about twice larger. Thus metallicity determinations need to be improved somewhat. SIM-Lite data, ground-based spectroscopy, and more advanced modeling of stellar atmospheres and evolution are absolutely crucial to establish an accurate calibration of the physical properties of old metal-poor halo stars, as well as determine the formation history of the oldest stars in our galaxy (e.g., Rollinde et al. 2008).

Once such calibrations have been determined, *single stars* can be used to determine the details of the assembly history of our Milky Way. Preliminary work on the formation of the thin disk has been done with Hipparcos data (e.g., Binney et al. 2000). However, ages for single stars will be about an order of magnitude less accurate than for binary stars[4]

At this point we would like to point out the unique abilities of a pointed SIM-Lite mission as compared to ESA's survey mission GAIA. While GAIA will obtain very accurate astrometry, spectroscopy and radial velocities for 10s to 100s of millions of stars, there are many types of objects that GAIA is not designed to handle very well. In a way, the "most interesting" stars fall in the not-GAIA category because they are rare, which means that they are typically found at large distances and/or relatively faint magnitudes. In those parts of parameter space, GAIA's performance falls well below the required accuracy. The most interesting stars are rare because they have a small space density (e.g., halo- or neutron stars) and/or because they are in a very short-lived phase of stellar evolution (e.g., supergiants, supernova precursors, Pop II Cepheids, long-period Pop I Cepheids, stars just before the core-helium flash, AGB stars in their mass-loss phase, and so forth). To go beyond observing and actually learn a great deal about these rare stars (mass, abundances, age etc.) we need to study them in detached eclipsing binary systems. And since double-lined detached eclipsing binaries amount to just 0.2% of the general stellar population, their average distances will be 22x (8x) larger than for *single* disk (halo) stars. Furthermore, it is often very desirable to learn in detail about the *individual stars. For such stars, SIM-Lite is far superior to GAIA* [see e.g. Unwin et al. (2008) and Benedict et al. (2009) for other examples of SIM-Lite's superior abilities in the field of stellar astrophysics].

For example, the highly accurate SIM-lite distances and ages of halo binaries will allow us to study the lifecycle of high-mass Pop III stars that have long ago enriched the interstellar medium. Possibly the only observable trace they left lies in the unique abundance pattern of stars that formed out of the ejecta created during the last stages of their life. Armed with SIM-Lite ages and proper motions supplemented by ground-based spectroscopy we might be able to identify groups of stars in 8-dimensional phase-space (3 space, 3 velocity, +age, + abundance pattern)

---

4 Masses of single stars can be determined from Newton's and Stefan-Boltzmann's laws: $M = g\, L / (4\, \pi\, G\, \sigma_{SB}\, T_{EFF}^4)$. All terms on the right-hand side are either fundamental constants or observables. The mass error ($\Delta M$) is mostly due to the gravity term (Olling 2003). We estimate that an improvement of the current temperature and gravity determinations (e.g., Valenti & Fischer 2005) by a factor of three allows for masses to be determined to ±5%. Then, Olling's (2003) approximate evolution rates indicate an age uncertainty of: $\Delta Age = [(\Delta L/L)^2 + (0.32 * Age * \Delta M)^2 ]^{1/2} / (-0.17 + 0.32 * M)$. For a 0.8 $M_\odot$ star aged 12 Gyr, the age error is 1.8 Gyr for 1% and 5% errors on L and M, respectively.

that originated from single Pop III stars.

To illustrate GAIA's strengths, the GAIA catalog will contain accurate phase space information for 10s of millions of stars within about a kpc from the Sun, while SIM-Lite supplements this data set for rare stars. Such data also allows one to determine the assembly history of the *disks of the Milky Way* (e.g., Freeman et al. 2002), as well as look for evidence of Galactic cannibalism: past (minor) merger events whose signatures may be recovered from a careful analysis of the phase-space data (e.g., McMillan & Binney 2008). In fact, such evidences have already been recovered from the Hipparcos data (e.g., Famaey et al. 2005; Helmi et al. 2006). Such 7D phase space information crucially constrains galaxy-formation models (e.g., De Lucia & Helmi 2008), and indeed even the formation redshifts of the earliest Pop III stars (Rollinde et al. 2008).

**GALAXIES:** If supplemented with ground-based spectroscopy and deep space-based imaging, accurate stellar ages can also be obtained for the rotational-parallax galaxies (M31 & M33). Thus, the formation histories of the Milky Way, M31 and M33 can be compared in great detail.

Such a giant step forwards in our understanding of stars and their spectra will also greatly impact studies of high-redshift galaxies that depend on measures of their integrated light (e.g., Lee et al. 2008) to, for example, determine their ages and star-formation histories (e.g., Panter et al. 2007). In this field, eclipsing binaries are also crucial for us to develop a firm understanding of young (pre-main-sequence) stars (Benedict et al. 2009). While the evolution of these objects is poorly understood (Hillenbrand & White 2005), they often dominate the light from high-z galaxies.

**THE UNIVERSE:** The Hubble constant is crucial for cosmology because the critical density depends only on $H_0$: $\rho_{CRIT} \equiv 3 H_0^2 / (8 \pi G)$. Thus, the accuracy with which we can determine even the simplest of cosmological parameter (the total density; $\Omega_{TOT}$) depends critically on the error on $H_0$: $\Delta\Omega_{TOT}/\Omega_{TOT} \geq 2 (\Delta H_0/H_0)$. Since the fate of the universe depends critically on the value of the critical density, accurate knowledge of the Hubble constant directly translate to the certainty with which we know $\Omega_{TOT}$. Put in another way, the CMB data alone cannot constrain separately the curvature and dark energy densities, while the addition of an independently determined Hubble constant results in strong constraints on the curvature of the universe.

One can determine the Hubble constant directly from the WMAP data *if one assumes* that the universe is flat ($\Omega_{TOT}\equiv 1$; e.g., Spergel et al. 2007) *and* that dark energy is in the form of a cosmological constant. The observed shape of the angular power spectrum depends mostly on the physical matter density ($\rho_{MATTER}$, e.g., in g/cm$^3$) which is determined by WMAP and $H_0$ (e.g. Hu 2005). Thus, for the case of an *assumed* flat universe and a cosmological constant, current data allows for the determination of the only unknown: $H_0$. Dunkeley et al. (2008) find $H_0 = 71.9 \pm 2.6$ km/s/Mpc,an error of 3.6%, while additional data reduces the error almost by a factor of two.

Likewise, the expected error on the equation of state of dark energy depends on the error on $H_0$ (Olling 2007). In fact, the error on the equation of state hardly declines if CMB data from the *Planck* mission is combined with the state-of-the art value & error of $H_0$ (Freedman et al. 2001) because in that case the contribution of the Hubble constant in the total error budget dominates, unless the error on $H_0$ is reduced to 1% or so (Olling 2007). As the quality of the additional data (supernovae, baryon-acoustic oscillations, galaxy clustering, etc.) improves, the importance of the contribution of $H_0$ declines. However, such inferences depend on the validity of "concordance cosmology," where for example, $H_0$ is derived from a number of data sets. Obviously, *one way to test the validity of concordance cosmology is to independently measure the value of the Hubble constant. Since cosmology current yields $H_0$ accurate to 2-4%, the independent determinations we propose here must have smaller errors still because future improvements* of *cosmology data sets will significantly reduce the concordance-errors for $H_0$.*

## DISTANCE MEASURES

All "intrinsic" methods for distance measurement need to be calibrated by a geometric method because they depend on some property of the object (e.g., the pulsation-luminosity-color relation for Cepheids) that depends on the specific physical state of the object. Typically we have poor a priori knowledge of this physical state, which may depend on numerous variables that are hard to measure such as: metallicity, the third parameter, α-element enhancement, age, μ-turbulent velocity, limb darkening, etc. In fact, many stars may have individual abundance patterns, possibly due to the particular SN debris from which they formed. This degree of individualness affects luminosity (Dotter et al. 2007) and may eventually limit the accuracy of standard candles. At the present time, we have no firm grasp of what the final limitations of the intrinsic methods are.

An important advantage of extrinsic techniques is that their results depend mostly on how much human ingenuity and technological prowess is applied to the problem, not on an intrinsic, and possibly unknowable, property of the target objects. Because "extrinsic techniques" are essentially geometric in nature, they may be more reliable than "intrinsic methods." On the other hand, some intrinsic technique may depend only slightly on the properties of "stars," while the required measurements are readily obtainable: such intrinsic method would be superior to geometric techniques. In fact, this has mostly been the situation in astrophysics to date, where parallax measurements are very difficult. One major task of SIM-Lite (and GAIA) is to ensure that geometric methods will have the upper hand, at least throughout most of the Milky Way. In addition, these astrometric missions *must* calibrate as many intrinsic methods as possible: after all, these methods move our distance horizon all the way to the end of the universe.

However, even "geometric" methods may have their pitfalls and need to be carefully calibrated (see Olling 2007 for a review). For example, the distances produced by two light-echo methods applied to SN1987A result in irreconcilable difference at the 10% level (Gould 2000). Also, because the extra-galactic water-maser method samples only 3 "lines of sight," this method is sensitive to systematic effects induced by non-circular motions: the allowed eccentric solutions (e=0.01 – 0.05; Humphreys et al. 2008) could be viable AGN-disk solutions (Armitage 2008), but they yield systematic distance errors twice the value of the eccentricity (2-10%;Olling 2007).

In conclusion, it is of crucial importance that most/all methods, intrinsic and extrinsic, are reliable in the statistical sense. *To be able to have confidence in the systematic errors -to move from model-based inferences to model-independent knowledge- many independent distance measures (intrinsic & extrinsic) need to be brought to the fore and scrutinized.* Only if such inter-comparisons are favorable, the measurements are laboratory-grade and can be reliably used to set the distance scale and/or determine stellar ages.

## SUMMARY & RECOMMENDATIONS:

**The next generation of micro-arcsec astrometric missions will contribute significantly to the fundamental goals for astrophysics as formulated by McKee & Taylor in the 2001 decadal report: "the fundamental goal ... is to understand how the universe ... galaxies [and] stars ... formed, how they evolved, and what their destiny will be."**

In order to do so, we recommend that US astronomy follows the recommendations of the previous two decadal committees in endorsing a mission like SIM-Lite that has substantial "wide-angle" capabilities (to measure absolute parallaxes and proper motions). We also recommend that US astrophysics maximize its involvement in the GAIA mission. Preferably via direct participation, and/or by facilitating the analysis of GAIA data via NASA/NSF grant programs. Given the complexity and abundance of the GAIA data, and the required ground-based data, it is crucial that US researchers can obtain funds to work on GAIA related science in the near future, but no later than the start of GAIA's operational phase. To determine accurate stellar ages also requires a moderate investment in ground-based high-resolution spectroscopy to obtain temperatures, metallicities and gravities of the target stars. An endorsement of the decadal committee of these projects would greatly help research in the arena of galactic archeology.

**The combination of these new data sets is needed to achieve our goal of being able to reconstruct all phases of the formation of the Milky Way and its nearby cousins. Specifically, we need: 1) SIM-Lite distances for thousands of rare halo stars, 2) large numbers of additional disk stars as obtained by GAIA, 3) improved analysis of high-resolution and high S/N ground-based spectra for 100s to 1,000s of stars, and 4) a SIM-Lite rotational parallax distance for M31 and M33 to geometrically establish the zero-point for virtually all distance indicators and to derive and compare the star-formation histories of these galaxies.**